\documentclass[12pt,epsf]{article}
\usepackage{graphicx, amsmath}

\begin{document}

\sloppy
\begin{flushright}{SIT-HEP/TM-28}
\end{flushright}
\vskip 1.5 truecm
\centerline{\large{\bf Topological curvatons}}
\vskip .75 truecm
\centerline{\bf Tomohiro Matsuda
\footnote{matsuda@sit.ac.jp}}
\vskip .4 truecm
\centerline {\it Laboratory of Physics, Saitama Institute of
 Technology,}
\centerline {\it Fusaiji, Okabe-machi, Saitama 369-0293, 
Japan}
\vskip 1. truecm
\makeatletter
\@addtoreset{equation}{section}
\def\theequation{\thesection.\arabic{equation}}
\makeatother
\vskip 1. truecm

\begin{abstract}
\hspace*{\parindent}
Topological defects may play the role of the curvatons.
We propose a new mechanism of generating density perturbations from
cosmological defects in inflationary models.
We show several examples in which defects play crucial role in
generating density perturbations.
\end{abstract}

\newpage
\section{Introduction}
\hspace*{\parindent}
In the standard scenario of the inflationary Universe, the observed
density perturbations are produced by a light inflaton field 
that rolls down its potential.
At the end of inflation, the inflaton oscillates about the minimum of
its potential and decays to reheat the Universe.
Adiabatic density perturbation is generated because the scale-invariant 
fluctuations
of the light inflaton field are different in different patches.
On the other hand, one may construct a model in which the ``light
field'' is not identified with the inflaton.
For example, the origin of the large-scale curvature perturbation in our
Universe might be induced by the late decay of a massive scalar field,
the  
curvaton\cite{curvaton_1, curvaton_2}.
The curvaton is assumed to be light during a period of cosmological
inflation so that it acquires scale-invariant fluctuations
 with the required spectrum.
After inflation, the curvaton starts to oscillate in a radiation
background. 
During this period, the energy density of the curvaton grows so that it 
accounts for the cosmological curvature perturbation when it decays.
The curvaton can generate the curvature perturbation after inflation if
its density becomes a significant fraction of the total energy
density of the Universe.
The curvaton paradigm has attracted a lot of attention because it was
thought to have an obvious advantage.
To be more precise, since the curvaton is independent of the inflaton
field,  
there was a hope\cite{curvaton_liberate}
 that the serious fine-tunings of the inflation models 
could be cured by the curvaton scenario, especially in models of
low-scale inflation\cite{low_inflation}. 
More recently, however, it has been suggested\cite{Lyth_constraint} that 
there is a strong bound for the Hubble parameter during inflation.
The bound obtained in ref.\cite{Lyth_constraint} seemed to be
crucial for inflationary models of a low inflation scale. 
Although the difficulty can be evaded\cite{matsuda_curvaton,
heavy_curvaton2} if there 
was an additional inflationary expansion (or a phase 
transition), it is still important to find another framework in which the
curvaton may accommodate low inflation scale\cite{matsuda_curvaton,
heavy_curvaton2}.
It should be helpful to note again that the original idea of the
curvaton is;
\begin{enumerate}
\item  The field(curvaton) other than inflaton acquires a perturbation
       with an almost scale-invariant spectrum and their density
       becomes a significant fraction of the total. 
\item  The curvaton decays into thermalized radiation so that the
       initial isocurvature density disappears.
\end{enumerate}
In the conventional curvaton scenario, generation of the curvature
perturbation is due to the oscillation of the curvaton that persists for
many Hubble times, which makes it possible for the curvaton to
dominate the energy density before it decays.\footnote{The idea
that a field other than inflaton acquires a perturbation 
with an almost scale-invariant spectrum can be used in different 
situations.
For example, one may construct a model\cite{decay-fluctuate} in which
the spatial fluctuations of the decay rate of the inflaton lead to the
fluctuations of the reheat temperature.
This scenario is different from the curvaton scenario in a sense that 
the reheating is still induced by the inflaton decay. 
In this case, unlike the curvaton scenario, the ``light field'' that
obtains scale-invariant fluctuations during inflation never dominates
the energy density of the Universe.
The scale-invariant fluctuations of the ``light field'' $\delta\phi$ is
transmitted to the curvature perturbation by the $\phi$-dependent
coupling constant and $\Gamma(\phi)$. 
The second paper in ref.\cite{decay-fluctuate} deals with superhorizon
fluctuations of the mass of a heavy particle.
These ideas may be similar to our scenarios of the monopoles, however the
decay mechanism is qualitatively different.} 
We agree with the original idea of the curvaton scenario,
however in any case one cannot simply ignore the possibility of
generating cosmological defects that might dominate the energy density
of the Universe.
Therefore, it is quite natural and important to investigate the  
possible scenario of ``topological curvatons'', in which the
domination by the ``curvaton'' is not due to the simple oscillation of the
curvaton field, but due to the evolution of the cosmological defects that
decay after they have become a significant fraction.
Along the lines of the above arguments, we propose a new
mechanism for generating density perturbations.
In the case that the cosmological defects dominate the energy
density and then decay to reheat (again) the Universe, the spatial
fluctuations of their ``nucleation rate'', ``nucleation time'' and
``decay rate'' may
lead to the fluctuations of the reheat temperature. 
The important point is that the scale-invariant
perturbation is due to a ``light field'' that obtains scale-invariant
fluctuations during inflation.
Therefore, the ``nucleation rate'', ``nucleation time'' or ``decay
rate'' of the topological curvatons must depend on the ``light field''.
Although our idea is mainly based on the original idea of the curvaton
scenario, there exists a crucial difference from the original.
In the original scenario of the curvaton, the
``light field'' is identified with the curvaton that dominates the
energy density of the Universe. 
In our model, however, the ``light field'' itself 
does not play the role of the curvaton, which may lead to some 
relaxation of the conditions in the original model. 
In this case, the mechanism that the scale-invariant fluctuations of the
``light field'' are transmitted to the topological curvatons is
important.

First, we should answer the question asking what the origin of
a scale-invariant perturbation is, in the case that cosmological
defects play the role of the curvaton.
In the scenario of the topological curvatons, we have two options.
The simplest idea is that a light field (or moduli) that obtains the
scale-invariant fluctuations during inflation leads to the
scale-invariant fluctuations of the ``decay rate'' of the
defects\cite{decay-fluctuate}.  
In this case, however, the fluctuations must remain until the defects
dominate and then decay to reheat the Universe.
This possibility is useful in the scenarios of the string overproduction
and the unstable domain walls, about which we will discuss later.
The important point is that the fluctuations of the decay rate cannot
survive until late epoch because it must be
damped when the ``light field'' starts oscillation.
In the scenarios of monopoles, we show that the scale-invariant
fluctuations in the annihilation rate lead to the fluctuations in the
initial (number) density when monopoles are frozen.

In this paper, we consider the case in which the scale-invariant
fluctuations of the vacuum expectation value of a light field 
$\sim \delta \phi/\phi$ leads to the similar fluctuations of a coupling
constant $\sim\delta g(\phi) /g(\phi)$.
This idea is quite similar to the one discussed in
ref.\cite{decay-fluctuate, Bauer_annihi}.
Therefore, we will assume that the
form of the field-dependent coupling is similar to the one that has
been used in the references. 
Please be sure that we are not considering any peculiar situation on
this point.  
Then, this may result in the fluctuations of ``nucleation
rate'',  ``nucleation time'', ``decay rate'' or ``annihilation rate''.
Obviously, the nucleation always occurs much earlier than the decay.
Therefore, the mechanisms that lead to fluctuations in the
initial density are useful if the required energy domination occurs
much later than the light-field oscillation.
Based on the above arguments, we consider a scenario in which the
scale-invariant 
fluctuations of the coupling constants lead to the similar fluctuations
of the initial density of the topological curvatons.
This scenario is discussed for the annihilating monopoles and the string
overproduction. 
It is important to note that the initial density fluctuations cannot
survive during the scaling epoch of the conventional strings and
domain walls.
However, we will show that in some cases the evolution before the scaling 
epoch may depend on the initial conditions.
Therefore, even in the cases of the strings or the domain walls, the
scale-invariant fluctuations of a ``light field'' may lead to the 
required density perturbation, if the defects decay safe
before it starts scaling.
We will discuss this issue in more detail for the string
overproduction. 
On the other hand, if the topological defects starts to dominate and
decay safe to ``reheat again'' the Universe soon after inflation,
the fluctuations in their decay rate $\delta \Gamma / \Gamma$ should
play significant role\cite{decay-fluctuate}. 
It should be noted that the latter idea might be conceptually different
from the original idea of the curvatons.

The advantage of our scenario is obvious.
In the conventional curvaton scenario, the scale of inflation is bounded
below by the requirements for the fluctuations and the
thermalization.\cite{Lyth_constraint}.
To avoid this difficulty, one should extend the original
scenario\cite{matsuda_curvaton}  so
that it includes additional phase transition that makes curvaton heavy
at later epoch.
To be precise, the curvaton must be light during inflation so that it
acquires the required scale-invariant fluctuations, while it must
become heavy in order to decay safe.
In our scenario, however, the ``light field'' that obtains
scale-invariant fluctuations during inflation does not decay to
``reheat again'' the Universe.
The secondary reheating that is required for the curvaton scenario
is induced by the topological curvatons. 
The important ingredient of our scenario is that {\bf the decay of the
topological curvaton is due to the characteristic
feature of the symmetry breaking or the characteristic evolution of the
resultant topological defects, which are independent of the
light field that obtains scale-invariant fluctuations during inflation.}

\section{Topological curvatons}
Although our scenario is rather different from the
conventional curvatons, we still think it is helpful to start
with the analogy with the original scenario.
The key ingredient of the original curvaton scenario is that the
scale-invariant fluctuations of 
the number density\footnote{Of course this expression is not precise.
However, for later discussions we note here that the evolution of the
oscillating curvaton looks quite similar to non-relativistic matter.}  
of the curvaton field, which have been generated at the time 
when 
curvaton field $\phi$ starts oscillation at the Hubble parameter
$H=H_{osc}\simeq  m_{\phi}$, remains until it dominates the Universe. 
Although the thermalization is induced by the late decay
of the curvaton, the initial fluctuations still remain and lead to the
fluctuation of the reheat temperature.

On the other hand, in the case that topological defects play the role of
the curvatons, it is not clear how the scale-invariant perturbations are
generated by a ``light field'' that is not identified with the
curvatons.
It is therefore quite useful to start with referring to the alternative
idea that 
has been advocated in ref.\cite{decay-fluctuate}.
The idea is quite simple and useful.
Since light fields ($\phi_i$) can acquire scale-invariant
fluctuations during inflation, $\phi_i$-dependent coupling constants
$g_j(\phi_i)$ may have the similar scale-invariant fluctuations.
These massless fields start oscillation when the Hubble constant becomes
smaller than their mass ($H_{osc} \simeq m_{\phi_i}$),
damping to their true vacuum expectation values.
Therefore, if the decay of the inflaton starts {\bf before} the damping,
the fluctuations of the coupling constants lead to
the fluctuations of the reheat temperature.  
In this case, $\delta T/T \sim \delta \Gamma/\Gamma \sim \delta
\phi_i/\phi_i$ is expected.(Numerical constants are neglected.)
Obviously, this scenario is useful for our scenario of the topological
curvatons, as we have commented in the previous section.

On the other hand, however, it is important to note again that 
the above mechanism cannot work in the scenario of the topological
curvatons if the domination starts at late epoch.
Therefore, in the case that the thermalization is induced by the
``late'' decay of the topological curvatons, the scale-invariant
fluctuations of the light 
fields must be converted into another type of fluctuations that will not
be damped before ``reheating''.
Obviously, the scenario that uses the fluctuations in the decay rate
is not suitable for the late-decay models.
In the conventional scenario of the curvatons, one can see that the 
fluctuation of the light curvaton is converted into the fluctuation 
of the oscillation, whose evolution is quite similar to the
number density of the non-relativistic matter.
Then the ratio of the curvaton density to the background
radiation increases with time, while the scale-invariant fluctuations
remain until the ``reheating''.
From the viewpoints that we have discussed above, it seems 
interesting to reconsider the curvaton scenario assuming that
the curvaton domination is not due to the conventional oscillation of
the curvaton 
field, but due to the evolution of the cosmological defects.
Our scenario is also interesting from the viewpoint of baryogenesis,
because decaying cosmological defects are sometimes considered as the
sources of the baryon-number asymmetry of the
Universe.\footnote{Inflationary models of low fundamental scale are 
discussed in ref.\cite{low_inflation, matsuda_nontach,
matsuda_defectinfla}. 
Scenarios of baryogenesis in such models are discussed in ref.
\cite{low_baryo, Defect-baryo-largeextra, 
Defect-baryo-4D}, where defects play distinguishable roles.}
Therefore, the most economical solution of our scenario is that the
defects are responsible for both the curvature perturbations and the
baryon-number asymmetry of the Universe.

It should be noted that the scenario of the topological
curvatons might be inevitable in a wide class of brane inflationary
models. 
To be precise, it has been discussed in ref.\cite{overproduction} that
the naive 
application of the Kibble mechanism may underestimate the initial density
of cosmic superstrings in brane inflationary models.
In this case, strings that are formed at
the end of brane inflation dominate the Universe soon
after inflation, and then they decay to ``reheat again'' the
Universe.\footnote{Brane defects such as monopoles, strings, domain
walls and 
Q-balls has been discussed in ref.\cite{overproduction, BraneQball,
matsuda_monopoles_and_walls, incidental, matsuda_coil}, where it was 
concluded that not only strings but also other defects should
appear after brane inflation.
Production of other topological defects such as monopoles and domain
walls is important and cannot be excluded in brane models.}
In this case, the strings that dominate the Universe after inflation are
the natural candidate of the topological curvatons.

\subsection{Fluctuations in initial number density (Monopoles)}
First, we will discuss about monopoles.
In this scenario, the required fluctuations are generated in the initial
number density of the monopoles.
To construct the successful scenario of the topological curvatons,
we need to discuss about a mechanism that triggers the efficient
annihilation of the monopoles\cite{Lang_and_Pi, mm-baryo}. 
This mechanism is often discussed in the context of the monopole
problem.   
Although there had been many possible solutions to the monopole
problem\cite{book_EU, blackhole}, it is commonly believed that the most
attractive solution would be the inflationary 
Universe scenario, in which the exponential expansion ensures the 
dilution of the preceding abundance of the unwanted monopoles.
However, other solutions are not useless because
in some cases monopoles could be produced {\bf after} inflation.
For example, apart from the conventional ``GUT monopoles'', monopole
is a generic topological defect that appears whenever the required
symmetry breaking occurs in the Universe.
Solutions of the monopole problem in which the inflationary expansion
is not assumed\cite{Lang_and_Pi, mm-baryo}
are therefore still important in some extended models.
In our scenario, since we are considering monopoles as the topological
curvatons, the monopoles must become a significant amount of the
Universe, and then decay safe by the mechanism of the efficient
annihilation\cite{Lang_and_Pi, mm-baryo}.

If the GUT transition is strongly first order, then the expected
relic monopole abundance is
\begin{equation}
\label{numberdensityofmono}
\frac{n_{M}}{s} \simeq \left[\frac{T_c}{M_p}ln\left(\frac{M_p^4}{T_c^4}
\right)\right]^3.
\end{equation}
where we have taken the grand unified transition temperature to be
$T_c \sim 10^{14}$GeV.
Here $n_M$ is the monopole number density, and $s$ is the entropy density
of the radiation.
Since the energy density of the radiation falls as $T^4$ while that of
the monopoles as $T^3$, the monopole to radiation energy density ratio
becomes $\sim 10^2$ by the time of the electroweak
transition\cite{mm-baryo}.
However, if there was a confining phase transition\cite{Lang_and_Pi,
mm-baryo}, monopoles will rapidly annihilate to release relativistic
particles that will rapidly thermalize to reheat the Universe.
Since we are considering a scenario of the topological
curvatons, we consider a model in which the transition occurs {\bf after}
monopoles have become a significant fraction of the Universe.

\underline{Efficient production and the origin of the scale-invariant perturbations}

Let us consider the ``worst'' scenario for the standard monopole
problem.
Here we consider the case in which the non-thermal process
leads to the efficient production of the monopoles after
inflation\cite{overproduction, PR}.  
In the case that $n_M/s > 10^{-10}$, Preskill\cite{preskill, earluUniv} found that
$n_M/s$ is reduced to about $10^{-10}$ by annihilations. 
In this case, the fluctuations in the annihilation cross section must
play important 
role in generating number density perturbations.
Recently, Bauer, Graesser and Salem(BGS) discussed in
ref.\cite{Bauer_annihi} that if the annihilation cross section receives
scale-invariant fluctuations, the required entropy perturbation is
produced at the freeze-out. 
In this case, if the particle comes to dominate the energy density of
the Universe and subsequently decay, this leads to the required 
adiabatic density perturbations.\footnote{Our model for monopoles as 
the topological curvatons is similar to the 
BGS mechanism, however there is a crucial difference in the decay
mechanism. 
As we have enphasezed in the introduction, the crucial ingredient
of the topological 
curvatons is that their decay is induced by the phase transition that 
is determined by the pattern of the symmetry breaking.}
In the BGS scenario, the number density of the heavy particle
$S$ is given by 
\begin{equation}
n_S\simeq \frac{T^3}{m_S M_{p}<\sigma v>},
\end{equation}
where $m_S$ and $<\sigma v>$ is the mass of the heavy particle and their
annihilation rate.
In our case, if monopoles are produced by an efficient process and
then decoupled from thermal equilibrium after
it has become non-relativistic, the number density of the monopoles at a
temperature $T$ after freeze-out is\cite{preskill}
\begin{equation}
n_m\simeq \frac{p-1}{A}\frac{m}{C M_{p}}\left(\frac{T}{m}\right)^{p-1}T^3,
\end{equation}
where we have followed ref.\cite{preskill} and used the equation
\begin{equation}
\frac{d n_m}{dt} = -\frac{A}{m^2}\left(\frac{m}{T}\right)^p n_m^2
-3H n_m
\end{equation}
and
\begin{equation}
H=\frac{T^2}{C M_p}.
\end{equation}
Here $A$ and $p$ are constants that characterize the annihilation
process. 
We have assumed that inflation itself cannot generate significant
density perturbations, which means that 
there is no significant fluctuation in the temperature at
the time of the freeze-out. 
We have also assumed that there is no fluctuation in the nucleation
time $t_c$, when monopoles are formed.
In the original BGS scenario, the decay temperature $T_{dec}$ is
determined by the decay rate $\Gamma$ and $<\sigma v>$.
On the other hand, in our case $T_{dec}$ is determined by 
the critical temperature $T_c'$ when the phase transition 
induces the efficient annihilation of the monopoles.
Therefore, in our case $T_{dom}$ and $T_{dec}$ in the BGS scenario
becomes\cite{preskill}
\begin{eqnarray}
T_{dom} &\simeq& 
\frac{p-1}{A}\frac{m^2}{C M_{p}}\left(\frac{T}{m}\right)^{p-1},\nonumber\\
T_{dec} &\simeq& T_{c}',
\end{eqnarray}
where $T_{dom}$ is the temperature when monopoles begin to dominate
the energy density.
Following the argument in ref.\cite{Bauer_annihi}, one can easily find the
energy density after monopole annihilation
\begin{equation}
\rho =\left(\frac{\rho(T_{dom})}{\rho(T_c')}\right)^{1/3}\rho_{rad},
\end{equation}
where 
\begin{eqnarray}
\rho(T_{dom}) &\simeq& T_{dom}^4 \propto A^{-4} m^{4(3-p)}\nonumber\\
\rho(T_c') &\simeq& T_c'^4 
\end{eqnarray}
In our case, the coupling to a light field can give rise to the
scale-invariant fluctuations in $A$ or $m$, which result in the
scale-invariant density perturbations
\begin{equation}
\frac{\delta \rho}{\rho} =
-\frac{4}{3}\frac{\delta A}{A}+\frac{4(3-p)}{3}\frac{\delta m}{m}.
\end{equation}
Once the field-dependent couplings are determined, one can easily
estimate the magnitude of the non-gaussian fluctuations.
As is discussed in ref.\cite{Bauer_annihi}, it is not
difficult to find a model in which
the non-gaussian fluctuations do not violate the observational
bound.\footnote{See ref.\cite{decay-fluctuate, Bauer_annihi} for more
details.} 
This highly model-dependent issue does not match the
purpose of this paper.

\underline{Second order phase transition and the origin of the
scale-invariant perturbations}

If the GUT transition is second order,
the expected relic monopole abundance is
\begin{equation}
\label{numberdensityofmono2ndorder}
\frac{n_{M}}{s} \simeq 10^2 \left(\frac{T_c}{M_p}\right)^3 \ll 10^{-10}.
\end{equation}
Therefore, the fluctuations in the annihilation cross section is not
important in this case\cite{preskill, earluUniv}, since the monopoles are
already ``frozen''.
In this case, the fluctuations in the critical temperature play
important role in generating fluctuations in $\delta n_M/n_M$.\footnote{
Be sure that the ``fluctuations in the critical temperature'' does not
mean the conventional fluctuations of the background temperature.
Here we are considering a peculiar situation where the background
temperature distribution is almost homogeneous, but the critical
temperature 
is different in different patches due to the scale-invariant 
fluctuations of the coupling constant.}
Here we assume that there is no vacuum-energy domination that induces
inflationary expansion or significant entropy production.\footnote{The
typical example would be thermal inflation\cite{thermal_inf}.}
The critical temperature $T_c$ for the monopole formation is 
\begin{equation}
T_c \simeq M_v \simeq e v, 
\end{equation}
where $M_v$ is the mass of the vector bosons, and $v$ is the vacuum
expectation value of the Higgs boson.
Obviously, the critical temperature 
depends on the coupling constant $e$ and the vacuum expectation value of
the Higgs boson $v$.
Assuming that the coefficients in the Higgs potential are
homogeneous, one may ignore the fluctuations in $v$.
In this case, $\delta e/e$ dominates the fluctuations.
Therefore, one can obtain 
\begin{equation}
\delta T_c/T_c \sim \delta e /e.
\end{equation}
 In this case, $\delta e/e$ leads to 
the spatial fluctuations of the initial number density of 
the monopoles(\ref{numberdensityofmono2ndorder}); 
\begin{equation}
\label{monoini}
\frac{\delta n_{M}}{n_{M}} \sim 3\frac{\delta e}{e}.
\end{equation}
The evolution of the non-relativistic monopoles  
is quite similar to the evolution of the oscillating curvatons.
In this case, the scale-invariant fluctuations that have been imprinted
on $\delta n_M/n_M$ are not damped during the evolution. 
Here we have assumed that the function of the coupling constant
$e(\phi)$ has the desired form, where $\phi$ is the light field that
obtains scale-invariant fluctuations during inflation.
As we have discussed above, it is not
difficult to find a model for $e(\phi)$ in which
the non-gaussian fluctuations do not violate the observational
bound. 
This highly model-dependent issue does not match the
purpose of this paper.

\underline{Entropy production}

If the phase transition is first order or there is an inflationary
expansion,  
one needs to consider an additional entropy production that is induced
by the ``reheating'' just after the phase transition.\footnote{This
``reheating'' is not due to the decay of the topological curvatons, but
is induced by the vacuum energy of the Higgs field.}
The entropy density increases by a factor of $\sim \Delta =(T_c/T_2)^3$,
where 
$T_c$ is the critical temperature and $T_2$ is the transition
temperature.
In this case, the fluctuations in the coupling constants may lead to the
fluctuations in the entropy increase factor $\delta \Delta/\Delta$
unless the 
fluctuations in $T_c$ cancels out the fluctuations in $T_2$.
Therefore, the two mechanisms (fluctuations in the entropy increase
factor and the topological curvatons) compete in this case.\footnote{The
generating mechanism of the scale-invariant perturbation from the
entropy increase  
factor $\Delta$ is a novel mechanism, although it may be regarded as a
variation of the conventional curvatons.
However, it is always difficult to calculate the transition temperature
$T_2$ for the strongly first order phase transitions.
On the other hand, we know a practical example of the entropy production.
In the scenario of thermal inflation\cite{thermal_inf}, $\Delta$ is simply
determined by the effective mass of the inflaton field near the origin.
In the case that the effective mass obtains the scale-invariant
fluctuations, which is of course due to a light field that obtains 
scale-invariant fluctuations during inflation, the resultant
fluctuations of $\Delta$ may play significant role.
We will discuss this mechanism in a separate
publication\cite{matsuda_future}.}

\underline{Baryogenesis}

As we have mentioned above, the most economical realization of our
scenario is to produce the baryon
number asymmetry from the decaying topological curvatons.
Dixit and Sher argued\cite{mm-baryo} that as the
Universe 
passes through the phase transition, the abundance of
monopoles will be depleted to acceptable levels, and at the same time
the process of annihilation can generate the observed baryon
asymmetry.  
The authors have considered an alternative version of the Langacker-Pi
scenario and discussed the scenario in which monopoles annihilate 
slightly after electroweak phase transition.
When they annihilate, their energy is released as relativistic
particles that rapidly thermalize.
At this time, combined with our idea of topological curvatons, the
fluctuations (\ref{monoini}) lead to
the fluctuations of the reheat temperature.
Then the resulting monopole abundance is reduced to $n_M/s \simeq
10^{-46}$, which is of course negligible today.
In this scenario, the monopole annihilation will significantly increase
the entropy of the Universe.
Therefore, unless there had been efficient mechanism of baryon number
production such as Affleck-Dine mechanism\cite{AD}, the observed baryon
asymmetry must be generated 
after the monopole annihilation.
The authors discussed\cite{mm-baryo} the generation of baryon number
asymmetry due to 
the annihilation of monopoles, which becomes
\begin{equation}
\frac{n_B}{s} \sim \epsilon \frac{T_R}{M_X},
\end{equation}
where $T_R \sim 10^3 GeV$ is the temperature of the radiation at the
time of the annihilation, $M_X \sim 10^{16} GeV$ is the mass of a heavy
particle, and $\epsilon$ is the average net baryon number produce in a
single annihilation.
Therefore, in the case that the fluctuations of the light fields 
lead to the fluctuations of the number density of the monopoles, our
scenario of topological curvatons works in this model.

Let us consider more generic constraint that bounds the mass scale of
the monopoles. 
If the phase transition is strongly first order, the ratio of the
monopole density to the background radiation is given by
\begin{equation}
\frac{\rho_M}{\rho_R} \simeq 10 \left(\frac{M_X}{M_p}\right)^3 M_X T^{-1}.
\end{equation}
In our scenario, monopoles must dominate the energy density of the
Universe before annihilation.
Therefore, the bound $\frac{\rho_M}{\rho_R} \gg 10^{-5}$ must be
fulfilled before the time of their annihilation.
Assuming that the annihilation occurs before the nucleosynthesis at 
$T\simeq 10 MeV$, we can obtain a bound 
\begin{equation}
M_X \gg 10^{11} GeV,
\end{equation}
which is obtained under stringent conditions.
If one considers non-thermal production of
monopoles\cite{overproduction, PR}, the above constraint can be
lowered. 

Obviously,  monopoles can play the required role of the topological
 curvatons. 
It is possible to construct models in which monopoles decay safe after
 they have dominated the energy density of the Universe.
The imprint on $n_M$ that has been induced by the scale-invariant
 fluctuations of the coupling 
 constant can survive during the evolution, and leads to the required
 density perturbation of the Universe.

\subsection{String overproduction}
The scenario of topological curvatons could be important in a wide class
of brane inflationary models.
It has been discussed in ref.\cite{overproduction} that
the naive 
application of the Kibble mechanism underestimates the initial density
of cosmic superstrings in brane inflationary models.
The authors argued that strings that are formed at the end of brane
inflation should dominate the Universe soon 
after inflation.
In this case, radiation becomes subdominant soon after inflation,
then strings decay to ``reheat again'' the Universe.
The mechanism of string overproduction is non-thermal, however their
density just before the ``reheating'' may depend on coupling constants.
Therefore, the strings produced by overproduction may play the role
of the topological curvatons. 
In this scenario, there could be three chances to transmit the
scale-invariant fluctuations of a ``light field'' to the strings.
\begin{enumerate}
\item The fluctuations are produced at the first stage of the
      overproduction.\\
      In this case, there are practical difficulties in estimating the
      density fluctuations.
      The mechanism of the overproduction would be highly non-linear,
      which means that the estimation of the density fluctuations 
      requires numerical calculation.
      This possibility is beyond the scope of this paper.
\item The first stage of the overproduction does not produce density
      fluctuations, however at the succeeding stage of the efficient
      loop production the fluctuations in the reconnection probability
      could lead to the density fluctuations of the small loops.\\
      We will investigate this possibility below.
\item Even if the string density is homogeneous for both the
      long strings and the string loops, fluctuations can be produced
      during their decay into radiation. 
      This possibility is similar to the mechanism that has been
      discussed in ref.\cite{decay-fluctuate}, despite the qualitative
      difference that in
      our case the reheating is induced by the decay of the topological
      defects. 
\end{enumerate}
In order to examine the above speculation of the topological curvatons
in the scenario of the string overproduction, let us assume that the
string 
density that is produced at the first stage of the string overproduction
is homogeneous, as we have discussed above.
First, we consider the second possibility that we have stated above.
The number density of the closed strings that are
produced succeedingly from the long strings may depend on the
reconnection probability $p$, which means that the density fluctuations of the
closed strings can be produced from $\delta p/p$, even if the string
density is homogeneous just after the first stage of the string
overproduction.  
In order to explain our idea, we consider the equations that
govern the energy density of the density of long strings ($\rho_L$) and
small loops ($\rho_l$)\cite{overproduction};
\begin{eqnarray}
\dot{\rho}_L &=& 
-2 H \rho_L -f \frac{\rho_L}{L}+\mu l \frac{d n_r}{dt}\nonumber\\
\dot{\rho}_l &=& 
-3 H \rho_l +f \frac{\rho_L}{L}-\mu l \frac{d n_r}{dt}
-\Gamma_s \rho_l,
\end{eqnarray}
where $ \frac{d n_r}{dt}$ and $\Gamma_s$ are the rate per unit volume
for loops to recombine with long strings and the string decay rate.
If the strings are not charged, $\Gamma_s$ is simply determined by the 
gravitational radiation.
Here $f(p)$ is a function of the reconnection probability of the strings, and
the typical length of the strings are denoted by $L$(long strings) and
$l$(string loops). 

In ref.\cite{overproduction}, it has been discussed that starting from
high density of the long strings and solving the above equations, the
Universe is soon dominated by the string loops but before long the
string loops decay into radiation.\footnote{See Fig.12
in ref.\cite{overproduction}. } 

Let us first examine the production of the closed string loops from 
the dense network of the long strings.
The above equations suggest that in the limit of the instant
production\footnote{In this case, one may assume that
$\dot{f}=\dot{L}=0$ during this period.}
the scale-invariant fluctuations in the reconnection probability $p$,
which is induced by a light field, lead to the scale-invariant density
perturbations of the closed strings.
To be precise, the long strings decay into small loops due to the second
term in r.h.s.($\sim -f(p) \frac{\rho_L}{L}$), which is proportional to
$f(p)$. 
Here $f(p)$ is the function of the reconnection probability,
which will go like $f \sim p^{1/2}$\cite{overproduction}.
Then, string loops dissipate their energy through radiation.
Therefore, the string loops that have been dominated the 
density of the Universe are soon converted into radiation that starts to
dominate the Universe after the ``reheating''.
Obviously, the reconnection probability of the cosmic strings does
depend on the effective coupling constants.
For fundamental strings, the reconnection probability is of order
$g_s^2$, then $f(p)$ becomes $f(p) \propto g_s$.
In other cases, since light fields or moduli of small
compactified dimensions determine the coupling constants in the
effective action, the scale-invariant fluctuations in the light
fields may lead to the scale-invariant density perturbations of the
string loops, $\delta\rho_l/\rho_l \propto \delta f/f$.\footnote{Even in
the standard inflationary scenario in which 
inflaton is 
responsible for the curvature perturbations, the perturbations that have
been produced by the inflaton reheating could be modified due to the
succeeding stage of the string domination.
In both cases the overproduction of cosmological defects could be used
to suggest the existence of tachyonic potential and brane inflation,
once the precise form of the effective action is determined.}
Therefore, in the case that $\Gamma_s$ is a homogeneous 
constant in the Universe, the required density perturbations can be
produced from $\delta \rho_l/\rho_l$.

The succeeding process of the reheating from string loops is quite
similar to the conventional reheating.
Therefore, the scenario proposed in ref.\cite{decay-fluctuate} is
successful in this case.
To be precise, the required perturbations can be produced by the
fluctuations in the decay rate $\delta \Gamma_s/\Gamma_s$, if $\Gamma_s$
is a desired function of a light field\cite{decay-fluctuate}.

The above two mechanisms ($\delta \rho_l/\rho_l$ and  $\delta
\Gamma/\Gamma$) work independently.

\section{Conclusions and Discussions}
\hspace*{\parindent}
In this paper, we have shown that the  new mechanism of generating density
perturbations can work in inflationary models.
We have considered monopoles and strings, and discussed about successful
scenarios. 
In the case of the monopoles, the spatial fluctuations of the initial
number density of the monopoles lead to the required fluctuations of
the Universe. 
Strings can play similar role if a dense string network is
produced by string overproduction.
Other defects may play the role of the topological curvatons, however
they are not so simple and require further study.
We will comment on the possibility and the difficulty of the domain
walls.

\underline{Domain walls}

Domain walls are produced when discrete symmetry is spontaneously
broken.
In the simplest case of $Z_2$-walls, it is known that the evolution of
the system of the domain walls becomes scale-invariant due to the
efficient 
annihilation and reconnection processes which suggests only one large 
wall exists per horizon.
In this case, the energy density of the walls is given by
\begin{equation}
\rho_w \sim \sigma t^{-1},
\end{equation}
where $\sigma$ is the tension of the domain wall.
Although the Universe could become wall-dominated at $t_*\sim
(G\sigma)^{-1}$, the wall-domination occurs (generally) after
scale-invariant evolution starts. 
This suggests that late-time behaviour of the system that could be
relevant for our scenario is not
sensitive to the choice of the initial state.

In a more complexified system of $Z_n$-walls, the domain walls
may become frustrated and their density decreases as\cite{kawasaki}
\begin{equation}
\rho_w \sim \sigma a^{-1},
\end{equation}
where $a$ is the scale factor.
The density of such domain walls will quickly dominate the Universe
before the system starts the scale-invariant evolution.
In this case, the initial density fluctuations of the domain walls 
could remain at the time of wall-domination and also affect 
the subsequent thermalization.
However, unlike the scenario of the string overproduction,
it is quite difficult to make a practical calculation
during this period.

On the other hand, 
one may expect that a coupling constant in the potential
of the domain walls depends on a light field $\phi$ that obtains a
scale-invariant fluctuation during inflation.
Then, in the case that the function $\sigma(\phi)$ has a desired form,
$\delta \phi/\phi$ may lead to the density fluctuations of the walls
\begin{equation}
\frac{\delta \rho_w}{\rho_w} \sim \frac{\delta\sigma}{\sigma} \sim
\frac{\delta \phi}{\phi}.
\end{equation}
This is a trivial example of the domain-wall curvatons.

Of course, it is possible to produce the required perturbations
from the scale-invariant fluctuations of the decay
rate of the walls $\delta\Gamma_w/\Gamma_w$, as we have
discussed for the strings\cite{decay-fluctuate}. 

The most important ingredient for the domain walls is the mechanism that
triggers the decay of the cosmological defects\cite{vilenkin_book}.
If the discrete symmetry is explicitly broken to a small extent,
the vacua are not degenerated and have slightly different energies.
The gap of the energies of the vacua leads to the gap of the pressure,
which finally leads to the instant collapse of the network of the domain
walls. 
In general, the magnitude of the explicit breaking term is determined by
an intentional higher-dimensional term, except for some peculiar
cases\cite{matsuda_wall}. 
Therefore, to obtain a successful cosmological scenario, a kind of
fine-tuning is required for the domain walls.

\section{Acknowledgment}
We wish to thank K.Shima for encouragement, and our colleagues in
Tokyo University for their kind hospitality.

\end{document}